УДК 539.188

# IMPROVED NEARSIDE–FARSIDE DECOMPOSITION
# OF ELASTIC SCATTERING AMPLITUDES

© 2004 r.   R. Anni, J. N. L. Connor*, C. Noli*

*Dipartimento di Fisica dell' Universita and Istituto
Nazionale di Fisica Nucleare, I73100 Lecce, Italy
*Department of Chemistry, University of Manchester,
Manchester M13 9PL, United Kingdom*

A recently proposed technique is described that provides improved nearsie-farside (NF) decompositions of elastic scattering amplitudes. The technique, involving a new resummation formula for Legendre partial wave series, reduces the importance of unphysical contributions to NF subamplitudes, which can appear in more conventional NF decompositions. Applications are made to a strong absorption model that arises in chemical and nuclear physics, as well as to a $^{16}O + ^{12}C$ optical potential at $E_{lab} = 132$ MeV.

## 1. INTRODUCTION

In nuclear, atomic and molecular collisions, an elastic differential cross section $\sigma(\theta)$, where $\theta$ is the scattering angle, is often characterized by a complicated interference pattern. In some cases, semiclassical methods [1] explain the scattering pattern as the interference between simpler, and slowly varying, subamplitudes. Ignoring the complication that, in some angular regions, uniform asymptotic techniques are often necessary, then the semiclassical subamplitudes arise mathematically from saddle points or poles which account physically for contributions from reflected, refracted or generalized diffracted semiclassical trajectories [2]. These subamplitudes can be conveniently grouped into two types: those arising from semiclassical trajectories which initially move in the same half plane as the detector (N or nearside trajectories) and those from the opposite half plane (F or farside trajectories).

Semiclassical methods are not always simple to apply and sometimes they have a limited range of applicability. In order to overcome these difficulties it is possible to apply a simple NF decomposition to the elastic scattering amplitude $f(\theta)$, that was proposed by Fuller [3] more than 25 years ago. Fuller's NF method uses only the quantum mechanical scattering matrix elements $S_l$, therefore bypassing the difficulties in using semiclassical methods. The aim of a NF method is to split the full scattering amplitude $f(\theta)$ into the sum of two subamplitudes, $f^{(-)}(\theta)$ and $f^{(+)}(\theta)$, describing the contributions from the N and F scattered particles, respectively.

The Fuller NF method was introduced to treat nucleus-nucleus scattering, where a long range Coulomb term is present in the interaction potential. Its starting point is the usual partial wave series (PWS) for $f(\theta)$

$$f(\theta) = \frac{1}{2ik}\sum_{l=0}^{\infty} a_l P_l(\cos\theta), \qquad (1)$$

where $k$ is the wavenumber, $P_l(\cos\theta)$ is the Legendre polynomial of degree $l$ and $a_l$ is given in terms of the scattering matrix element $S_l$ by:

$$a_l = (2l+1)(S_l - 1). \qquad (2)$$

We recall, in fact, that the PWS in (1), considered as a distribution, is convergent also if $S_l$ is asymptotically Coulombic [4].

In Section 2, we outline the Fuller NF method and discuss some of its limitations. Section 3 describes an improved NF method using resummation techniques that reduces the importance of unphysical contributions to the NF subamplitudes. Results from our improved NF method for the elastic scattering of two collision systems are presented in Section 4. Our conclusions are in Section 5.

## 2. FULLER NF METHOD
## AND ITS LIMITATIONS

The Fuller NF decomposition is realized by splitting $P_l(\cos\theta)$, considered as a standing angular wave, into traveling angular wave components

$$P_l(\cos\theta) = Q_l^{(-)}(\cos\theta) + Q_l^{(+)}(\cos\theta), \qquad (3)$$

where (for $\theta \neq 0, \pi$)

$$Q_l^{(\mp)}(\cos\theta) = \frac{1}{2}\left[P_l(\cos\theta) \pm \frac{2i}{\pi}Q_l(\cos\theta)\right], \qquad (4)$$

* Electronic address: J.N.L.Connor@Manchester.ac.uk

with $Q_l(\cos\theta)$ the Legendre function of the second kind of degree $l$.

Inserting (3) into (1), splits $f(\theta)$ into the sum of two subamplitudes $f^{(\mp)}(\theta)$. For $l\sin\theta \gg 1$, the $Q_l^{(\mp)}(\cos\theta)$ behave as

$$Q_l^{(\mp)}(\cos\theta) \sim \sqrt{\frac{1}{2\pi\lambda\sin\theta}}\exp\left[\mp i\left(\lambda\theta - \frac{\pi}{4}\right)\right], \quad (5)$$

with $\lambda = l + 1/2$. This asymptotic behavior suggests that the $f^{(\mp)}(\theta)$ should correspond to NF trajectories respectively appearing in the complete semiclassical decomposition of $f(\theta)$ (Ref. [1], p. 121).

The NF decomposition is in general less satisfactory than the full semiclassical one. For example, if two or more N or F semiclassical trajectories contribute to the same $\theta$, interference effects may appear in the N or F cross sections. The Fuller NF decomposition has, however, the merit of being simple and, although inspired by the semiclassical theories, it uses only quantities calculated within the exact quantum mechanical treatment. The NF method therefore bypasses problems associated with the applicability and validity of semiclassical theories.

The physical meaning attributed to the $f^{(\mp)}(\theta)$ is implicitly based on the (unproven) hypothesis that it is possible to perform on the PWS, written in terms of the $Q_l^{(\mp)}(\cos\theta)$, the same manipulations that are used in deriving the complete semiclassical decomposition of $f(\theta)$. These manipulations are path deformations in $\lambda$ of the integrals into which (1) can be transformed, using either the Poisson summation formula or the Watson transformation. The consequences of these path deformations depend on the properties of the terms in the PWS when they are continued to real or complex values of $\lambda$ from the initial half integer $\lambda$ values. The splitting of $P_l(\cos\theta)$ into $Q_l^{(\mp)}(\cos\theta)$ modifies these properties and can cause the appearance of unphysical contributions in the $f^{(\mp)}(\theta)$ which cancel out in $f(\theta)$.

In spite of these possible limitations, the Fuller NF decomposition is widely used, as demonstrated by the fact that the ISI Web of Science reports about 140 citations since 1981 to the original Fuller work. In many nucleus-nucleus scattering cases [5, 6], the Fuller method effectively decomposes $f(\theta)$ into simpler subamplitudes, which are free from the unphysical contributions that can arise from the above mathematical difficulties. However for a few examples, the NF subamplitudes can be directly compared with the corresponding semiclassical results and it is found that the Fuller and semiclassical decompositions predict different results. One classic example is pure Coulomb scattering. For repulsive Coulomb potentials only a N contribution is expected semiclassically ([1], p. 56), whereas the Fuller NF decomposition yields also a F contribution [3]. Another important example is the angular distribution for a strong absorption model (SAM) with a two parameter ($\Lambda$ and $\Delta$) symmetric $S$-matrix element and Fermi-like form factors [7]

$$S_l \equiv S(\lambda) =$$
$$= \left[1 + \exp\left(\frac{\Lambda - \lambda}{\Delta}\right)\right]^{-1} + \left[1 + \exp\left(\frac{\Lambda + \lambda}{\Delta}\right)\right]^{-1}, \quad (6)$$

with $\lambda = l + 1/2$. For a fixed value of the cut-off parameter $\Lambda$ and for a sufficiently large value of the diffuseness parameter $\Delta$, the Fuller NF cross sections agree with the semiclassical results only up to a certain value of $\theta$, which decreases with increasing $\Delta$.

Fortunately, the Fuller NF subamplitudes contain information that allows one to recognize the unphysical nature of the undesired contributions. Suppose $f^{(+)}(\theta)$, or $f^{(-)}(\theta)$, contains a single contribution from a stationary phase point at $\lambda(\theta)$. Then the derivative with respect to $\theta$ of the phase of $f^{(+)}(\theta)$, or $f^{(-)}(\theta)$, is equal to $\lambda(\theta)$, or $-\lambda(\theta)$ respectively. Following Fuller we will call this derivative the *Local Angular Momentum* (LAM) for the N (or F) subamplitude; it depends on $\theta$ ([1], p. 57). Only for certain generalized diffracted trajectories is the LAM expected to be constant, equal to the angular momentum of the incoming particle responsible for the diffraction. In the semiclassical regime, this constant value is expected to be large. Because of this, if we observe that in a certain $\theta$ range LAM $\approx 0$, this can be considered the signature of the unphysical nature of the N or F subamplitudes in that range of $\theta$. This occurs for the LAM of the Fuller Coulomb F subamplitude, and for the NF subamplitudes of the SAM in the angular region where the Fuller NF cross sections differ from the semiclassical results. In both cases this decoupling of $\theta$ from LAM suggests the unphysical nature of the subamplitudes. Thus an analysis of the LAM can avoid misleading interpretations of cross sections obtained from the Fuller NF decomposition. However the problem of obtaining more satisfactory NF decompositions remains open.

## 3. IMPROVED NF METHOD USING RESUMMATION THEORY

A possible solution to the problem was proposed by Hatchell [7], who used a modified NF decomposition. The modifications consisted of, first, writing $f(\theta)$ in the resumed form ($\theta \neq 0$)

$$f(\theta) = \frac{1}{2ik}\frac{1}{(1 - \cos\theta)^r}\sum_{l=0}^{\infty} a_l^{(r)} P_l(\cos\theta), \quad (7)$$

$r = 1, 2, \ldots$, and, second, using a different splitting for the Legendre polynomials into travelling waves.

The use of the resummed form (7) for $f(\theta)$ was originally proposed [8] by Yennie, Ravenhall, and Wilson (YRW) to speed up the convergence of the PWS for high-energy electron-nucleus scattering. Equation (7) is an exact resummation formula, of order $r$, which is derived from the recurrence relation for Legendre polynomials. The YRW resummation formula can be derived by iterating $r$ times, starting from $a_l^{(0)} = a_l$, the resummation identity

$$\sum_{l=0}^{\infty} a_l^{(i-1)} P_l(\cos\theta) = \frac{1}{1-\cos\theta} \sum_{l=0}^{\infty} a_l^{(i)} P_l(\cos\theta), \quad (8)$$

$i = 1, 2, \ldots$, where

$$a_l^{(i)} = -\frac{l}{2l-1} a_{l-1}^{(i-1)} + a_l^{(i-1)} - \frac{l+1}{2l+3} a_{l+1}^{(i-1)}, \quad (9)$$

with $a_{-1}^{(i-1)} = 0$.

Note that $f(\theta)$ is independent of $r$, unlike the Fuller NF subamplitudes which do depend on the value of $r$ used. This is a consequence of the property $lQ_{l-1}(\cos\theta) \longrightarrow 1$ as $l \longrightarrow 0$ [9]. In the Hatchell approach, the dependence on $r$ arises because the functions used in place of the $Q_l(\cos\theta)$ obey a (inhomogeneous) recurrence relation different from that for $Q_l(\cos\theta)$. It is worthwhile to note that (7), for $r \geq 1$, makes irrelevant the 1 appearing in the term $(S_l - 1)$ in (2) for $\theta > 0$. The linear transformation from $a_l^{(i-1)}$ to $a_l^{(i)}$ drops contributions to $a_l^{(i)}$ from any constant additive term in $a_l^{(i-1)}$. Furthermore for $r \geq 1$, (7) produces a convergent PWS even when $S_l$ is asymptotically Coulombic [10].

Using his method, Hatchell has shown [7] that the unphysical contributions to the SAM NF cross sections systematically decrease on increasing $r$. More recently [11], it was shown that, using the Fuller $Q_l^{(\mp)}(\cos\theta)$ functions in (7), gives even better results. The superiority of the $Q_l^{(\mp)}(\cos\theta)$ functions seems to be connected with the greater rapidity with which the $Q_l^{(\mp)}(\cos\theta)$ approach their asymptotic behavior (5) [12, 13], compared to the functions used by Hatchell to split $P_l(\cos\theta)$ into traveling angular waves.

The success of using (7) before applying the NF splitting (3), (4) of $P_l(\cos\theta)$ depends on the properties of the $a_l^{(r)}$. For the SAM, the contributions form low $l$ values rapidly decrease [11] with increasing $r$. As a result, the most important contributions move to higher values of $l$, where a semiclassical description is physically more reasonable.

However, in some cases, (7) acts in the opposite direction, by enhancing the undesired unphysical contributions to the NF subamplitudes. We have found that this happens, for example, for pure Coulomb scattering, for scattering by an impenetrable sphere, and for the SAM [14] when the cross section is calculated at an angle $\pi - \theta$, using the property $P_l[\cos(\pi - \theta)] = (-1)^l P_l(\cos\theta)$.

One possible solution to this puzzle was recently found by introducing an improved NF method [15, 16], based, first, in dropping the 1 from the term $(S_l - 1)$ in the PWS for $f(\theta)$, and, second, in considering (8) only as a particular case of a modified resummation identity [17]

$$\sum_{l=0}^{\infty} a_l^{(i-1)} P_l(\cos\theta) = \frac{1}{\alpha_i + \beta_i \cos\theta} \sum_{l=0}^{\infty} a_l^{(i)} P_l(\cos\theta), \quad (10)$$

with $\alpha_i + \beta_i \cos\theta \neq 0$ and

$$a_l^{(i)} = \beta_i \frac{l}{2l-1} a_{l-1}^{(i-1)} + \alpha_i a_l^{(i-1)} + \beta_i \frac{l+1}{2l+3} a_{l+1}^{(i-1)}. \quad (11)$$

For $\alpha_i$, $\beta_i \neq 0$, the r.h.s. of (10) depends, apart a renormalization factor, only on the ratio $\beta_i / \alpha_i$. Thus, without loss of generality, we can assume $\alpha_i = 1$ for all $i$. By iterating (10) $r$ times, we can write $f(\theta)$ in the modified resummed form

$$f(\theta) = \frac{1}{2ik} \left( \prod_{i=1}^{r} \frac{1}{1 + \beta_i \cos\theta} \right) \sum_{l=0}^{\infty} a_l^{(r)} P_l(\cos\theta), \quad (12)$$

$r = 1, 2, \ldots$ The identity (12) is the key result for improving the NF decomposition. The YRW resummation formula (7) is obtained with $\beta_1 = \beta_2 = \ldots = \beta_r = -1$.

The utility of dropping the 1 from the term $(S_l - 1)$ in the PWS is suggested by the fact that

$$f_\delta(\theta) = \frac{i}{2k} \sum_{l=0}^{\infty} (2l+1) P_l(\cos\theta) = \frac{i}{k} \delta(1 - \cos\theta), \quad (13)$$

where $\delta$ indicates the Dirac $\delta$-function. We then obtain [18]

$$\begin{aligned} f_\delta^{(\mp)}(\theta) &= \frac{i}{2k} \sum_{l=0}^{\infty} (2l+1) Q_l^{(\mp)}(\cos\theta) \\ &= \pm \frac{1}{2\pi k(1 - \cos\theta)}. \end{aligned} \quad (14)$$

This means that the dropped term does not contribute, for $\theta \neq 0$, to the full scattering amplitude $f(\theta)$, but it gives a contribution to the NF subamplitudes $f^{(\mp)}(\theta)$. This contribution does not depend in any way on the properties of $S_l$, and, because of this, it is difficult to think that it can have a physical meaning.

The utility of using the resummation identity (10), rather than the YRW (8) used by Hatchell, is suggested by the fact that this resummation identity with $\vec{\beta} \equiv \beta_1 = \ldots = \beta_r$ is a particular case of a more general one [17], which uses a basis set of reduced rotation matrix elements; this gives the amplitude for more general scattering processes than those described by (1). For these general PWS, a Fuller-like NF decomposition can be introduced [19–21] which allows the scattering amplitude to be split into NF subamplitudes. In some cases, the NF cross sections contain unexpected (unphysical) oscillations [17], which are enhanced if the generalization of (7) is used, but which disappear for an appropriate choice of the β-parameter in the generalization of (12).

Similarly, the unphysical contribution to the NF subamplitudes of the SAM [14] when the scattering amplitude is calculated at $\pi - \theta$ decrease by increasing the order of the resummation with the choice $\beta \equiv \beta_1 = \ldots = \beta_r = 1$. On the contrary, using the Fuller NF decomposition in the YRW (7), for which $\beta \equiv \beta_1 = \ldots = \beta_r = -1$, the unphysical contributions become larger on increasing the resummation order.

These results suggest that the modified resummed form (12) be used to diminish unphysical contaminations to the NF subamplitudes. To do this, we must give a practical rule to fix the values of the $\beta_i$ parameters. In Refs. [14, 17], it was proposed to select the value of $\beta \equiv \beta_1 = \ldots = \beta_r$ so that $(1 + \beta\cos\theta)^{-r}$ *approximately mimics the shape of the angular distribution*. The shape of the cross section can however be very different from that given by $(1 + \beta\cos\theta)^{-r}$. It is therefore desirable to use a different recipe, based possibly on a simple rule. The quantitative recipe proposed in [15, 16] is inspired by the observation that the modified resummation formulas produce a more physical NF decomposition by reducing the contribution from the low $l$ values in the resummed PWS. This suggests we select the $\beta_1, \beta_2, \ldots, \beta_r$ in $r$ repeated applications of (10), so as to eliminate as many low $l$ contributions as possible from the final PWS in (12). The transformation from $\{a_l^{(i-1)}\}$ to $\{a_l^{(i)}\}$ is linear tridiagonal, with coefficients linear in $\beta_i$, which means application of $r$ successive resummations allows one to equate to zero the leading $r$ coefficients $a_l^{(r)}$, with $l = 0, 1, \ldots, r-1$, by solving a system of $r$ equations of degree $r$ in the parameters $\beta_1, \beta_2, \ldots, \beta_r$. The resummation defined in this way was named [15, 16] an *improved resummation of order r*.

It is straightforward to show that the improved resummation of order 1 is obtained by choosing

$$\beta_1 = -3a_0/a_1, \tag{15}$$

while the improved resummation of order 2 is given by

$$\beta_{1,2} = (B \pm \sqrt{B^2 - 4A})/2, \tag{16}$$

with $A$ and $B$ solutions of the linear equations

$$\begin{cases} \left(\dfrac{1}{3}a_0 + \dfrac{2}{15}a_2\right)A + \dfrac{1}{3}a_1 B = -a_0, \\[2mm] \left(\dfrac{3}{5}a_1 + \dfrac{6}{35}a_3\right)A + \left(a_0 + \dfrac{2}{5}a_2\right)B = -a_1. \end{cases} \tag{17}$$

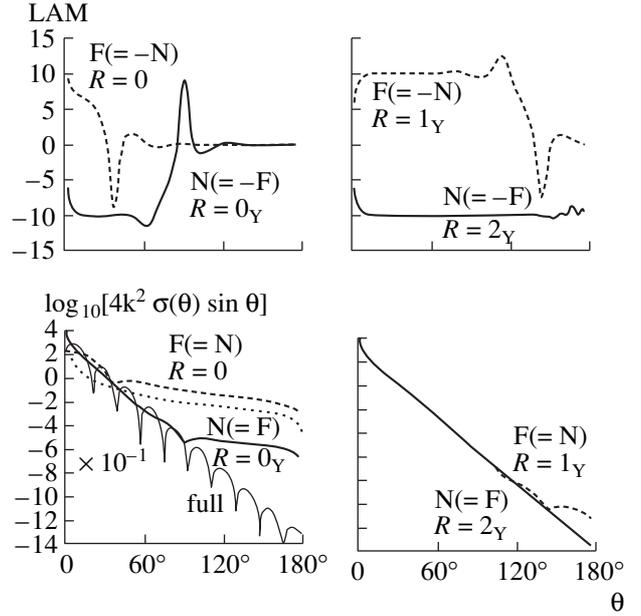

**Fig. 1.** Strong absorption model N (continuous lines) and F (dashed lines) cross sections (lower panels) and LAM (upper panels) calculated using the $R = 0$, $0_Y$, $1_Y$, $2_Y$ NF decompositions. The thin curves show the full cross section. The thin dotted curve shows the F (=N) cross section (displaced downward by one unit) for the unphysical amplitude $f_\delta^{(+)}(\theta)$.

Higher order improved resummations require the solution of more complicated systems of equations.

In all cases analyzed using $r \leq 2$ [15, 16, 18, 22], the improved NF method considerably reduced the width of the angular regions in which the Fuller NF cross sections exhibit unphysical behavior. These analyses used $S_l$ from simple parameterizations as well as from some optical potentials currently employed to describe light heavy-ion scattering.

## 4. RESULTS FOR ELASTIC SCATTERING

We show below our results [15] for two of these examples. The first example is a SAM (Fig. 1). The SAM is a standard test case for NF methods [7, 11, 13], which can be applied to atomic, molecular and nuclear collisions in the energy range meV to GeV, since it does not depend on the energy. The second example is the $^{16}O + ^{12}C$ collision, at $E_{lab} = 132$ MeV, using the WS1 optical potential of Ref. [23] (Fig. 2). The cross section for this potential exhibits many intriguing properties of light heavy-ion cross sections in the energy range around 10 MeV/A, where A is the mass number of the projectile.

The improved NF method drops the 1 that ensures the convergence of the original PWS for the SAM. For the optical potential case, the Coulombic behavior of $S_l$ for large values of $l$, makes (2) not convergent in the usual sense, with or without the 1. In both the cases

LIP, fm

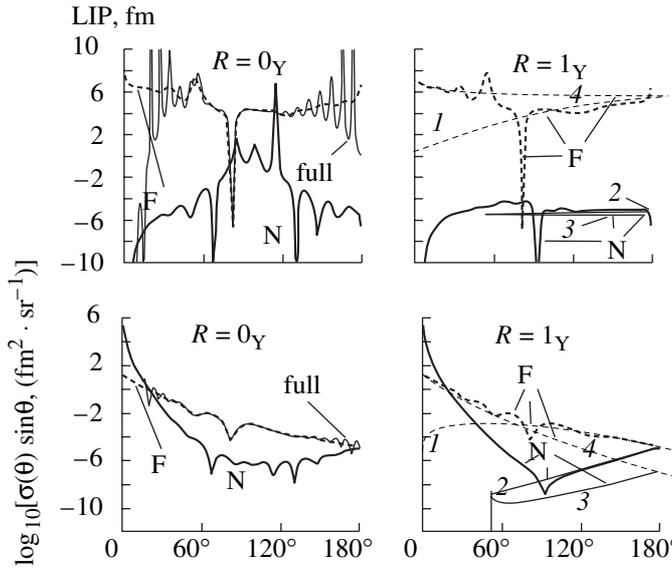

**Fig. 2.** Optical potential model N (continuous lines) and F (dashed lines) cross sections (lower panels) and LIP (upper panels) calculated using the $R = 0_Y$ and $R = 1_Y$ NF decompositions. The thin curves show the cross section and LIP obtained using the full quantum amplitude in the left panels, and the N (continuous) and F (dashed) cross sections and LIP using classical mechanics in the right panels. The indices in the right panels identify the curves corresponding to different branches of the classical deflection function.

presented here, the PWS defining the NF subamplitudes are only defined in a distributional sense, and some technique must be used to make these series convergent in the usual sense. As shown in Ref. [15, 18], an additional extended YRW resummation, of appropriate order, can be used to ensure (and also speed up) the convergence of these series.

The calculations were therefore performed applying: *first*, an improved resummation of order $r = 0, 1, 2$, with $r = 0$ meaning no resummation, *second*, the Fuller NF splitting (3), (4), and *third* a YRW resummation of the NF subamplitudes using the extension of (7) to the linear combination (4) of integer degree Legendre functions of the first and second kinds [9]. This latter resummation ensures the convergence of the final NF PWS. The results obtained from these three steps will be indicated by the notation $R = 0_Y, 1_Y, 2_Y$.

### A. Strong absorption model

For the SAM we show the results obtained using the parameters $\Lambda = 10$ and $\Delta = 2$. In Fig. 1 we have plotted the dimensionless quantity $4k^2\sigma(\theta)\sin\theta$ since the corresponding NF semiclassical quantities are expected to have a pure exponential slope [7]. Furthermore, because the $S_l$ are real, $f(\theta)$ has a constant phase (and its phase derivative is of no interest), and the $f^{(\mp)}(\theta)$ have the same moduli but opposite phases. Thus we need only show the N, or F, LAM and similarly for the cross sections.

Figure 1 also shows the results obtained on introducing the NF decomposition directly into (1) and without dropping the 1 in (2). The NF subamplitudes obtained in this way are rapidly convergent and no final extended YRW resummation is needed. These results are indicated by the notation $R = 0$.

For the original Fuller NF method, $R = 0$, and for the case $R = 1_Y$, we have plotted the F cross sections and LAMs (dashed curves); for the cases $R = 0_Y$ and $R = 2_Y$ the N quantities (continuous curves) are displayed.

Figure 1 shows that for $R = 0$ the unphysical contributions dominate the $F$ (=N) cross section over most of the angular range. The expected exponential behavior is not present in the F cross section curve and the F (=–N) LAM $\approx 0$ for $\theta \gtrsim 60°$. However, at smaller angles, oscillations in the F LAM curve indicate that another contribution is present which interferes with the unphysical one. This behavior does not support the conjecture that the F LAM of this other contribution has the semiclassical value $\Lambda$. The major part of the unphysical contribution, which dominates the F subamplitude, is the F subamplitude $f_\delta^{(+)}(\theta)$ (14) of the $\delta$-function contribution (13) to the amplitude obtained by dropping $S_l$ in the term $(S_l - 1)$ in (2). The cross section for $f_\delta^{(+)}(\theta)$ is shown, downward shifted by one vertical unit, by the thin dotted curve in Fig. 1.

The $R = 0_Y$ method provides more satisfactory results, which are rather good at forward angles. With the exclusion of a small region around $\theta = 0°$, where (5) does not hold, the N (=–F) LAM agrees closely with the expected semiclassical value of $-\Lambda$ up to $\theta \approx 50°$, and the N (=F) cross section curve follows the expected exponential behavior. For $\theta \gtrsim 120°$, the N cross section is still dominated by an unphysical contribution. At intermediate angles, $50° \lesssim \theta \lesssim 120°$, interference oscillations appear both in the N cross section and in the N LAM curve. It is interesting to note that the LAM is more sensitive to interference effects than is the cross section. Also, in the interference region, one cannot attach the meaning of a *local angular momentum* to the subamplitude phase derivative. In our case, in this interference region, the N LAM curve oscillates around the expected semiclassical value of $-\Lambda$ in the region, $50° \lesssim \theta \lesssim 80°$, where the true semicalssical component dominates the N subamplitude, and around the unphysical value of 0 at larger angles.

The effectiveness of the improved resummation procedure is evident in the right panels of Fig. 1. Using the $R = 1_Y$ method (for which $\beta_1 = -0.800$) the F (=–N) LAM and the F (=N) cross section are in agreement with the semiclassical results up to $\theta \approx 120°$. For $R = 2_Y$ (which has $\beta_{1,2} = -0.879 \pm 0.076i$), the agreement cov-ers almost the whole angular range. The small irregular oscillations appearing at large $\theta$ for the N LAM curve, with $R = 2_Y$, probably arise from the precision limitations (64 bit floating point representation) of the calculations.

## B. Optical potential

Figure 2 shows our results for the optical potential. In the upper panels we display LAM/k, which we call the *Local Impact Parameter* (LIP), and in the lower panels a plot of $\log_{10}[\sigma(\theta)\sin\theta]$. The left panels show the results for the usual Fuller NF decomposition, $R = 0_Y$. The thin continuous lines, in the left panels, show the cross section and LIP for the full amplitude.

The behavior of the NF LIP curves is mostly simpler than that of the full amplitude LIP. At $\theta \approx 90°$, the $R = 0_Y$ N LIP curve oscillates around 0, indicating the possible dominance of an unphysical contribution. This contribution also appears to be responsible for oscillations in the N LIP curve around other values (different from 0), and for oscillations in the N cross section for $\theta \gtrsim 30°$. These oscillations are absent in the N curves in the right panels, where the results for $R = 1_Y$ are shown $(\beta_1 = -0.999 - 0.099i)$. Both the N cross section and N LIP curves for $R = 1_Y$ are considerably simpler than those obtained using $R = 0_Y$, while the F curves are essentially the same; an exception is the less oscillatory F LIP for $\theta \gtrsim 120°$. This indicates that, apart from $\theta \gtrsim 120°$, the unphysical contribution for $R = 0_Y$ has a modulus which is much smaller than that of the F semiclassical subamplitude. We have also applied the improved resummation $R = 2_Y$. The results are practically the same as those obtained using $R = 1_Y$ and are not shown.

The cleaning by the $R = 1_Y$ procedure of the original $R = 0_Y$ NF subamplitudes is impressive and allows a clear identification, in the NF cross sections at $\theta \gtrsim 120°$, of the dominance of semiclassical trajectories refracted from the internal part of the nuclear interaction. In the right panels of Fig. 2, this interpretation is confirmed by the agreement, for $\theta \gtrsim 120°$, between the NF curves and the classical mechanical results (thin lines *1* and *2*) corresponding to impact parameters smaller than the nuclear rainbow one. The thin lines show, in the upper panel, different NF branches of the impact parameter and their dependence on $\theta$ (with appropriate signs) using only the real part of the complete interaction (Ref. [22], Fig. 1). In the lower panel we show the classical contributions to the cross section from these branches, in which we have included in the usual simple way ([1], p. 49) the absorptive effects of the imaginary part of the optical potential.

## 5. CONCLUSIONS

Our new resummation NF procedure clearly improves the original Fuller NF decomposition, as is evident in the examples presented here. On the one hand, our results confirm the importance of NF decompositions for gaining insight into the properties of the subamplitudes responsible for complicated structures in cross sections. On the other hand, they confirm the empirical origin of NF decompositions and suggest caution in the interpretation of results obtained from NF techniques. However, different NF decompositions can be used to check what parts of the resulting NF subamplitudes are independent of the particular technique used. Only properties stable with respect to different NF decompositions, can be considered as manifestations of some physical phenomenon. In addition, we have shown that it is desirable to investigate the behavior of the LAM. This quantity is more sensitive to interference effects than are the NF cross sections, and a null value (or an oscillatory behavior around zero) of the LAM in a certain angular range may indicate the dominance of an unphysical contribution.

Further examples of the effectiveness of the improved NF technique, here briefly described, can be found in Refs. [16, 18, 22].

## ACKNOWLEDGMENTS

Support of this research by a PRIN MIUR (I) research grant, the Engineering and Physical Sciences Research Council (UK) and INTAS (EU) is gratefully acknowledged.

## REFERENCES

1. *Brink D.M.* Semi-Classical Methods for Nucleus-Nucleus Scattering. Cambridge: University Press, 1985.

2. *Nussenzveig H.M.* Diffraction Effects in Semiclassical Scattering. Cambridge: University Press, 1992.

3. *Fuller R.C.* // Phys. Rev. C. 1975. V. 12. P. 1561.

4. *Taylor J.R.* // Nuovo Cimento. B. 1974. V. 23. P. 313.

5. *Hussein M.S., McVoy K.W.* // Progr. Part. Phys. 1984. V. 12. P. 103.

6. *Brandan M.E., Satchler G.R.* // Phys. Rep. 1997. V. 285. P. 143.

7. *Hatchell P.J.* // Phys. Rev. 1989. V. 40. P. 27.

8. *Yennie D.R., Ravenhall D.G., Wilson R.N.* // Phys. Rev. 1954. V. 95. P. 500.

9. *Anni R., Renna L.* // Lett. Nuovo Cimento. 1981. V. 30. P. 229.

10. *Goodmanson D.M., Taylor J.R.* // J. Math. Phys. 1980. V. 21. P. 2202.

11. *Hollifield J.J., Connor J.N.L.* // Molec. Phys. 1999. V. 97. P. 293.

12. *McCabe P., Connor J.N.L.* // J. Chem. Phys. 1995. V. 104. P. 2297.

13. *Hollifield J.J., Connor J.N.L.* // Phys. Rev. A. 1999. V. 59. P. 1694.

14. *Noli C., Connor J.N.L.* // Rus. J. Phys. Chem., Suppl. 1. 2002. V. 76. P. S77.

15. *Anni R., Connor J.N.L., Noli C.* // nucl-th. 2001. / 0111060.

16. *Anni R.* nucl-th. 2002. / 0208036.

17. *Whiteley T.W.J., Noli C., Connor J.N.L.* // J. Phys. Chem. A. 2001. V. 105. P. 2792.

18. *Anni R., Connor J.N.L., Noli C.* // Phys. Rev. C. 2002. V. 66. 044610; nucl-th. 2002. / 0207059.

19. *Dobbyn A.J., McCabe P., Connor J.N.L., Castillo J.F.* // Phys. Chem. Chem. Phys. 1999. V. 1. P. 1115.

20. *Sokolovski D., Connor J.N.L.* // Chem. Phys. Lett. 1999. V. 305. P. 238.

21. *McCabe P., Connor J.N.L., Sokolovski D.* // J. Chem. Phys. 2001. V. 114. P. 5194.

22. *Anni R.* // Eur. Phys. J. A. 2002. V. 15. P. 361.

23. *Ogloblin A.A., Glukhov Y.A., Trzaska W.H. et al.* // Phys. Rev. C. 2000. V. 62. P. 044601.